\documentclass{ifacconf}

\usepackage{graphicx} 
\usepackage{epsfig}   
\usepackage{mathptmx} 
\usepackage{times}    
\usepackage{amsmath}  
\usepackage{amssymb}  
\usepackage{natbib}   
\usepackage{anyfontsize}
\usepackage{calrsfs}
\usepackage{todonotes}
\usepackage{enumitem}
\usepackage{algorithmic}
\usepackage{setspace}
\usepackage{color,soul}
\DeclareMathAlphabet{\pazocal}{OMS}{zplm}{m}{n}

\newcommand{\norm}[1]{\left\lVert#1\right\rVert}

\makeatletter
\def\endfigure{\end@float}
\def\endtable{\end@float}
\makeatother

\let\ifacconfcaptionwidth\captionwidth
\usepackage[caption=false]{subfig}
\let\captionwidth\ifacconfcaptionwidth

\begin{document}
\begin{frontmatter}

\title{Linear Time-Periodic System Identification with Grouped Atomic Norm Regularization} 

\thanks[footnoteinfo]{This work is supported by the Swiss National Science Foundation under grant
no.: 200021\_178890, and the Swiss Competence Center for Energy Research SCCER FEEB\&D of the Swiss Innovation Agency Innosuisse.}
\thanks[copyright]{\copyright\ 2020 the authors. This work has been accepted to IFAC for publication under a Creative Commons Licence CC-BY-NC-ND.}

\author[First]{Mingzhou Yin}
\author[First]{Andrea Iannelli}
\author[First]{Mohammad Khosravi}
\author[First]{Anilkumar Parsi}
\author[First]{Roy S. Smith}

\address[First]{Automatic Control Laboratory, ETH Z\"{u}rich, Switzerland \\ (e-mail: \{myin,iannelli,khosravm,aparsi,rsmith\}@control.ee.ethz.ch).}

\begin{abstract}                
This paper proposes a new methodology in linear time-periodic (LTP) system identification. In contrast to previous methods that totally separate dynamics at different tag times for identification, the method focuses on imposing appropriate structural constraints on the linear time-invariant (LTI) reformulation of LTP systems. This method adopts a periodically-switched truncated infinite impulse response model for LTP systems, where the structural constraints are interpreted as the requirement to place the poles of the non-truncated models at the same locations for all sub-models. This constraint is imposed by combining the atomic norm regularization framework for LTI systems with the group lasso technique in regression. As a result, the estimated system is both uniform and low-order, which is hard to achieve with other existing estimators. Monte Carlo simulation shows that the grouped atomic norm method does not only show better results compared to other regularized methods, but also outperforms the subspace identification method under high noise levels in terms of model fitting.
\end{abstract}


\end{frontmatter}

\section{Introduction}

Linear time-periodic (LTP) systems are an important type of system that sees a wide range of applications in rotating machinery (\cite{Allen_2011}), aerospace (\cite{Shin_2005,Wood_2018}), power systems (\cite{2000}), process control (\cite{Budman_2013}), etc., to model periodicity in dynamics, scheduling parameters, and operating trajectories. This paper focuses on the identification of LTP systems. This topic has received considerable attention due to its close connections with identification of linear time-varying systems (\cite{Liu_1997}), linear parameter-varying systems (\cite{Felici_2007}), and nonlinear systems along limit cycles (\cite{Allen_2009}).

In general, any identification scheme for linear time-invariant (LTI) systems is applicable to LTP systems by application of the lifting technique by \cite{Bittanti_2000}. However, such methods often fail to encode characteristics of lifted systems, such as the causality constraint that prevents future inputs in a period affect previous outputs. Thus, the identified lifted system is not guaranteed to be realizable as its LTP form. Specific extensions to LTP systems are also available. An extension of the subspace identification method by \cite{Verhaegen_1995} is widely applied (e.g. \cite{Wood_2018,Liu_1997,Felici_2007}). Its frequency-domain counterpart has been recently proposed by \cite{Uyanik_2019}. This method suffers from the model order selection problem at high noise level, especially for LTP systems that require the selected system order to be consistent at all tag times, as discussed in \cite{Wood_2018}. Harmonic transfer function (HTF) coefficients are identified in \cite{Louarroudi_2012} and \cite{Yin_2009} using least-square methods. This approach leads to high-order non-parametric models. Non-convex optimization methods are used to directly identify state-space models in \cite{Goos_2016}, with the drawback that globally optimal estimates are not guaranteed to be achieved.

As in the above methods, it is desired to decompose or lift LTP systems to structured LTI models and extend existing LTI system identification frameworks to them. However, as pointed out in \cite{Bittanti_2000}, the key issue in this process is that the parameters in these structured models have strong correlations since they come from the same dynamic system. This correlation is not investigated in existing LTI-based identification frameworks.

On another note, regularized optimization has reported positive results in linear system identification recently (see e.g. \cite{Chen_2012,Pillonetto_2016,Smith_2014,Shah_2012}) after its success in statistics and machine learning. The underlying idea of regularization techniques is to separate the objectives of maximizing data adherence and incorporating prior knowledge on system structure by employing distinct terms. This makes it possible to use simple models to depict complex system structures. Typical LTI system structures investigated include stability, continuity, and low complexity. In particular to LTP systems, this framework enables us to treat the previously discussed correlation issue with parameter regularization.

This paper focuses on identifying a low-McMillan-degree single-input and single-output (SISO) model. This low-order assumption is both practical for common physical systems and useful for various control design problems. To circumvent the hard model complexity selection problem, the regularized method is used with a general high-order model and a complexity penalization term. The most widely-used complexity penalization is the rank of the Hankel operator and its convex surrogate, the Hankel nuclear norm proposed by \cite{Fazel_2001}. To improve stability with finite data lengths and computational scalability, the atomic norm was proposed in \cite{Shah_2012} which uses $l_1$-regularization to select a finite number of order-revealing atomic dynamics.

For LTP systems, there is an additional constraint in the low-order estimation: the identified model should have a consistent system order throughout the period. This requirement is very practical yet hard to achieve with existing methods. For the rest of the paper, this requirement is referred as uniformity.

The contribution of this paper is to propose a methodology to identify uniform low-order models for LTP systems. The proposed method extends atomic norm identification for LTI systems and applies group lasso regularization to impose the additional constraints needed for periodic models. Group lasso was proposed in \cite{Yuan_2006} to solve the grouped factor selection problem in regression and then used in numerous optimization problems including applications in identification of switched systems (\cite{Ohlsson_2013}), dynamic networks (\cite{Chiuso_2012}), and non-linear systems with heterogeneous data (\cite{Pan_2018}). In our method, parameters are grouped based on the fact that the LTI sub-models should always select the same atomic dynamics with the same poles. A case study and Monte-Carlo simulation show that our proposed method is not only effective in estimating uniform low-order LTP models, but also superior to existing methods in terms of model fitting under high noise levels.

The remainder of the paper is organized as follows. Section~\ref{sec:form} defines the LTP system to be identified. Section~\ref{sec:LTI} introduces LTI reformulation of LTP systems and formulates a least squares problem for identification. Section~\ref{sec:reg} extends regularization techniques to LTP system and proposes a uniform and low-rank regularizer. The validity and effectiveness of the proposed method are illustrated in Section~\ref{sec:sim0} by simulation. Section~\ref{sec:con} concludes the paper.

\section{Problem Statement}
\label{sec:form}

Consider a discrete-time SISO LTP system that follows the minimal state-space realization
\begin{equation}
\begin{cases}
x(t+1)&=\ A(t) x(t)+B(t) u(t)\\
\hfil y(t)&=\ C(t) x(t)
\label{eq:sys}
\end{cases},
\end{equation}
where $x(t) \in \mathbb{R}^{n_x}$, $u(t) \in \mathbb{R}$, and $y(t) \in \mathbb{R}$ are the states, input, and output respectively. The time-varying matrices $A(t)=A(t+P),B(t)=B(t+P),C(t)=C(t+P)$ are periodic state-space matrices of appropriate dimensions, and $P$ is the period. The stability of LTP systems can be assessed by the spectral radius of the monodromy matrix $\Psi_{A,\tau}=A(\tau-1)A(\tau-2)\cdots A(\tau-P)$. \cite{Bittanti_1986} proved that the eigenvalues of $\Psi_{A,\tau}$ are independent of $\tau$ and that the system is stable iff the spectral radius $\rho(\Psi_{A,\tau})<1$.

In the remainder of the paper, the following system identification problem is considered:
\begin{description}
	\item[\textbf{Given:}] sequences of the true input $u(t)$ and the noise contaminated output of the system (\ref{eq:sys}): $z(t)=y(t)+w(t)$, where $w(t)$ is the unknown noise, for $t = 1,2,\cdots,nP$, where $n$ is the number of periods observed.
	\item[\textbf{Assumptions:}] 1) the period length $P$ is known; 2) the system is stable, i.e., $\rho(\Psi_{A,\tau})<1$; 3) the system is of low McMillan degree, i.e., $n_x \ll n$; 4) the noise is Gaussian with $w(t) \sim N(0,\sigma^2)$; 
	\item[\textbf{Objective:}] estimate a uniform low-order model of the system (\ref{eq:sys}).
\end{description}

\section{LTI Reformulation of LTP Systems}
\label{sec:LTI}

In this section, methods to reformulate LTP systems as structured LTI models are reviewed. Based on the reformulation, a least squares problem is formulated to identify switched finite impulse response (FIR) models of LTP systems without structural constraints.

\subsection{Lifting and Switching}

Lifting and switching are two main reformulations of LTP systems to apply LTI methods. The lifting method converts the LTP system to an ordinary LTI system of $P$-times larger input and output dimensions and $P$-times slower. The state dimension remains the same. Due to its natural connection to the subspace identification formulation, the method is usually used to extend the subspace identification method. Readers are referred to \cite{Bittanti_2000,Wood_2018} for more details about the lifting method and its application in subspace identification.

\begin{figure}[htbp]
\centerline{\includegraphics[width=0.85\linewidth]{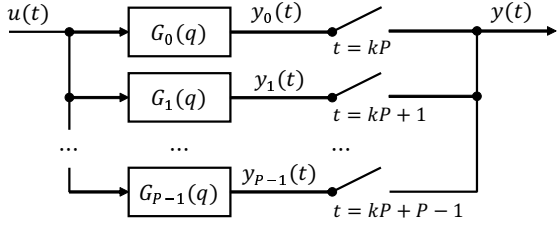}}
\caption{Illustration of switching reformulation of LTP systems.}
\label{fig:1}
\end{figure}

In the switching method, the LTP system is reformulated as a switched LTI system with $P$ switches. In detail, the system (\ref{eq:sys}) is expressed with the following input-output model
\begin{equation}
y(t) = \sum^\infty_{i=1} g^t_i u(t-i),
\label{eq:smod}
\end{equation}
where
\begin{equation}
g^t_{i}=C(t)A(t-1)A(t-2)\cdots A(t-i+1)B(t-i),
\label{eq:imp}
\end{equation}
where the superscript $t$ denotes the current tag time, and the subscript $i$ denotes the time difference between the input and the output. Since the dynamics are periodic, $g^t_i$ is also $P$-periodic with respect to $t$. For a fixed $t$, $\{g^t_i\}$ formulates a valid infinite impulse response (IIR) model of a LTI system as
\begin{equation}
G_\tau(q)=\sum^\infty_{i=1} g^\tau_i q^{-i} = C(\tau)(q^P I_{n_x}-\Psi_{A,\tau})^{-1}\pazocal{B}_\tau(q),
\end{equation}
where
\begin{equation}
\pazocal{B}_\tau(q) = \sum_{i=0}^{P-1}A(\tau-1)A(\tau-2)\cdots A(\tau+i-P+1)B(\tau+i)\cdot q^i,
\end{equation}
$q$ is the forward time-shift operator, $\tau = 1,2,\cdots,P$. The  models $G_\tau(q)$ will be called sub-models in the following. Thus, we define a periodically switched LTI model of the LTP system as
\begin{equation}
y(t) = y_\tau(t),\ t = kP+\tau,
\end{equation}
where $y_\tau(t) = G_\tau(q) u(t)$. See Fig.~\ref{fig:1} for a diagrammatic illustration. Note that the dynamics of each switch $G_\tau(q)$ have exactly the same poles, which are the solutions to $f_{\Psi}(q^P)=0$, where $f_{\Psi}(x)$ is the characteristic polynomial of $\Psi_{A,\tau}$. The solutions are independent of $\tau$ because they are the $P$-th roots of the eigenvalues of the monodromy matrix, which are independent of $\tau$. Therefore, for a uniform LTP system (\ref{eq:sys}), the poles in each sub-model are exactly the same. The system order of the switched system is then $P\cdot n_x$. This reformulation has been used to estimate HTFs in \cite{Yin_2009}.

Comparing both methods, lifting constructs a system that is $P^2$-times larger than the original system, whereas switching decomposes the system into $P$ sub-systems of the same size. The additional parameters induced by redundant dimensions in lifting are constrained by causality constraints on the lifted systems, i.e., future inputs cannot affect previous outputs. These causality constraints are in general difficult to enforce in identification except in the subspace framework. This also induces computational scalability issues. Switching, conversely, preserves the input-output dimensions of the LTP system at the expense of augmented system orders. However, this problem can be alleviated by using regularization techniques, where the computation does not scale with the system order, as a sufficiently high-order model is needed to start with anyway.

\subsection{The Least Squares Problem for the Switched model}

With the switched model, a least squares problem can be formulated to estimate the impulse response coefficients that minimizes the following quadratic objective function.
\begin{equation}
V_{LS}(\mathbf{g}|u(t),z(t))=\sum_{\tau=1}^{P} \sum_{k=0}^{n-1} \left[z(kP+\tau) - \sum^N_{i=1} g^{\tau}_i u(kP+\tau-i)\right]^2,
\label{eq:LS}
\end{equation}
where
\begin{equation}
\mathbf{g}=\begin{bmatrix}
g^1_1 & g^2_1 & \cdots & g^P_1\\
g^1_2 & g^2_2 & \cdots & g^P_2\\
\vdots & \vdots & \ddots & \cdots\\
g^1_N & g^2_N & \cdots & g^P_N\\
\end{bmatrix}\in \mathbb{R}^{N \times P}
\end{equation}
gathers the parameters in all sub-models. Note that here the IIR models are truncated to $N$ terms and hence become FIR models.

However, this unregularized problem does not enforce the requirement that the identified system should be uniform and low order. In addition, the impulse responses need to be stable. Attempts to formulate these requirements in a regularization term are discussed in the following section.

\section{Low-Order Regularization of LTP systems}
\label{sec:reg}

In its general form, regularization techniques investigate the following optimization problem.
\begin{equation}
\underset{\mathbf{g}}{\text{minimize}}\ V(\mathbf{g})+\gamma\cdot J(\mathbf{g}),
\end{equation}
where $V(g)$ is the loss function that evaluates data adherence, such as $V_{LS}(\mathbf{g})$. The term $J(g)$ is the regularizer that encodes prior knowledge of the model, and $\gamma\geq 0$ is a scalar weighting factor that balances two objectives. 

To the best of our knowledge, regularized techniques have yet to be applied to LTP systems. This section investigates the extension of low-order regularizers to LTP systems. We first discuss the common rank regularizer to see why it is not suitable for LTP system identification, and introduce an approximately equivalent alternative - the atomic norm. It is shown that a grouped version of the atomic norm can effectively regularize the estimator to be uniform and low-order.

\subsection{Rank regularization}

The most common low-order regularization is based on the fact that the rank of the extended observability and controllability matrices gives the McMillan degree of the system. Different matrices have been constructed that reveal this rank for regularization. In the switched IIR model (\ref{eq:smod}), the Hankel operator on the impulse responses,
\begin{equation}
\pazocal{H}(\mathbf{g}^t) =
\begin{bmatrix}
g^t_1&g^t_2&\cdots&g^t_{N-m+1}\\
g^t_2&g^t_3&\cdots&g^t_{N-m+2}\\
\vdots&\vdots&\ddots&\vdots\\
g^t_m&g^t_{m+1}&\cdots&g^t_{N}\\
\end{bmatrix},
\end{equation}
is commonly used as the rank-revealing matrix, where $\mathbf{g}^t=[g^t_1\ g^t_2\ \cdots\ g^t_N]^T$ is the $t$-th column of $\mathbf{g}$. Since the rank function is highly non-convex, its best convex surrogate, the nuclear norm, is applied in optimization for tractability as in \cite{Smith_2014}. Thus, we have the following nuclear norm regularizer
\begin{equation}
J_N(\mathbf{g}) = \sum_{\tau=1}^{P} \beta_\tau \norm{\pazocal{H}(\mathbf{g}^\tau)}_*,
\label{eq:nuc}
\end{equation}
where $\norm{\cdot}_*$ denotes the nuclear norm, which is the sum of the singular values, and $\beta$ is the weighting vector of sub-model complexity.

However, in addition to its general issue of stability (\cite{Pillonetto_2016}) and scalability (\cite{Shah_2012}), the Hankel nuclear norm regularizer fails to provide an explicit expression for the system order. This makes it hard to tune different sub-models to the same order, not to mention the requirement of the same pole locations. In fact, as demonstrated in Section~\ref{sec:sim}, this regularizer often cannot regularize the sub-systems to any given order despite fine tuning of the weighting vector $\beta$.

\subsection{Atomic norm regularization}
\label{sec:atom}

The atomic norm regularization was proposed in order to overcome the stability and scalability issues of the Hankel nuclear norm. The underlying idea is to replace the search for a rank-revealing system matrix with the search for an order-revealing decomposition of the system. As proposed in \cite{Shah_2012}, consider a set of stable first-order systems
\begin{equation}
\pazocal{A}=\left\{a_w(q)=\left.\frac{1-|w|^2}{q-w}\ \right\rvert \ w \in \mathbb{D}\right\},
\end{equation}
where $\mathbb{D}$ is the open unit disk in the complex plane. The elements in the set are dubbed ``atoms" and are normalized to have a Hankel nuclear norm of 1. This selection of atoms guarantees the stability of the estimated system. As such, assuming the sub-models have no repeated poles, i.e., the monodromy matrix $\Psi_{A,\tau}$ is diagonalizable, the sub-models can be decomposed as linear combinations of atoms by performing partial fraction expansions of the transfer functions,
\begin{equation}
G_\tau(q)=\sum_{w \in \mathbb{D}} c^\tau_w\cdot a_{w}(q)\approx \sum_{k=1}^{n_p} c^\tau_k\cdot a_{w_k}(q):=\mathbf{c}_\tau^T \mathbf{a}(q),
\label{eq:pfe}
\end{equation}
where the infinite atom set is approximated by fine gridding $\{w_k\}$ with an atom vector $\mathbf{a}(q)=[a_{w_1}(q)\ a_{w_2}(q)\ \cdots \ a_{w_{np}}(q)]^T$. The vector $\mathbf{c}_\tau=[c^\tau_1\ c^\tau_2\ \cdots \ c^\tau_{n_p}]^T$ denotes the corresponding coefficients, and $n_p$ is the number of atoms in the grid. Note that the coefficients $c_k^\tau$ are complex. When repeated poles exist, atoms of higher order can be included.

In this case, the McMillan order of the system is equal to the cardinality of $\mathbf{c}_\tau$. It is well-known that the best convex surrogate for the cardinality function is the $l_1$-norm. The technique to use $l_1$-norm to promote sparsity is often known as lasso. The atomic norm of the system $G_\tau(q)$ is defined as
\begin{equation}
\norm{G_\tau(q)}_\pazocal{A}=\norm{\mathbf{c}_\tau}_1.
\end{equation}
It was shown in \cite{Shah_2012} that the atomic norm is a good approximation to the Hankel nuclear norm.

As we are dealing with real-valued systems, the partial fraction expansion (\ref{eq:pfe}) includes either real poles or conjugate pairs of poles. The coefficients corresponding to the conjugate pairs are also required to be conjugate to one another. This imposes additional constraints on $\mathbf{c}_\tau$ as
\begin{equation}
c^\tau_k = \text{conj}(c^\tau_l),\forall w_k=\text{conj}(w_l),\tau=1,2,\cdots,P.
\label{eq:st}
\end{equation}

To apply the atomic norm regularization to the switched model, the partial fraction expansion (\ref{eq:pfe}) is rewritten in terms of impulse responses as 
\begin{equation}
\mathbf{g} = \mathbf{g}^a \mathbf{c},
\end{equation}
where $\mathbf{g}^a = [\mathbf{g}^a_1\ \mathbf{g}^a_2\ \cdots\ \mathbf{g}^a_{n_p}] \in \mathbb{R}^{N \times n_p}$, $\mathbf{g}^a_k$ is the $N$-truncated impulse response of $a_{w_k}(q)$, and $\mathbf{c} = [\mathbf{c}_1\ \mathbf{c}_2\ \cdots\ \mathbf{c}_P]\in \mathbb{C}^{n_p \times P}$. Note that $\mathbf{g}^a$ is a constant matrix that can be pre-computed. Thus, we have the following atomic-norm regularized optimization problem.
\begin{equation}
\begin{matrix}
\underset{\mathbf{c}}{\text{minimize}}&\ V_{LS}(\mathbf{g}^a \mathbf{c})+\gamma \cdot J_A(\mathbf{c}),\\
\text{subject to}&\ \ (\ref{eq:st})
\end{matrix}
\label{eq:atom}
\end{equation}
where
\begin{equation}
J_A(\mathbf{c}) = \sum_{\tau=1}^{P} \beta_\tau \norm{\mathbf{c}_\tau}_1.
\label{eq:areg}
\end{equation}

Since the first-order atoms are smooth and stable, the estimated system satisfies the smoothness and stability constraints. In addition, problem (\ref{eq:atom}) is a quadratic programming (QP) problem, which has much better scalability compared to the semidefinite programming (SDP) problem induced by the nuclear norm regularization. However, since each sub-model is still separately regularized, the uniformity requirements are still not guaranteed. Crucially, in contrast to the nuclear norm regularizer, we now have the pole location information from the estimated parameters $\mathbf{c}$. In the next subsection, this information will be used to propose a uniform regularizer that guarantees the same pole locations for each sub-model.

\subsection{Grouped Atomic Norm Regularization}
\label{sec:group}

The basic idea to modify the previous LTI-based atomic norm regularizer (\ref{eq:areg}) to satisfy LTP requirements is to connect the same atom at different tag time. In general terms, the same atom needs to be either included in all or excluded from all of the sub-model dynamics. To do this, we first examine the structure of the parameter matrix $\mathbf{c}$. If the $(i,j)$-th element in $\mathbf{c}$ is non-zero, it means the the sub-model $j$ has a pole at $w_i$ and vice versa. Therefore, in addition to the sparsity requirement induced by the low-order assumption, each row of $\mathbf{c}$ also needs to be either all zero or all non-zero. This requirement coincides with the concept of grouping in group lasso.

Group lasso, also known as sum-of-norms, is an extension of lasso or $l_1$-norm regularization to enforce sparsity on groups of parameters rather than isolated parameters. Consider a set of grouped parameters $\{\theta_i\}, \theta_i \in \mathbb{R}^{m_i}, i=1,2,\cdots,M$. The group lasso regularizer is
\begin{equation}
J_G(\{\theta_i\}) = \sum_{i=1}^M \norm{\theta_i}_2.
\end{equation}
Here, $l_2$-norms are used to relax the sparsity constraint inside each group, and the sparsity-promoting function reduces to summation since the $l_2$-norms are always non-negative. In this way, sparsity is enforced on the group $l_2$-norms: when the $l_2$ norm is regularized to zero, all parameters in the group are zero; when the $l_2$ norm is non-zero, all the parameters are usually non-zero. So consistent sparsity is promoted inside each group. In particular, for the parameter matrix $\mathbf{c}$, each row is collected as a group. So the following grouped atomic norm regularizer is proposed
\begin{equation}
J_{GA}(\mathbf{c}) = \sum_{k=1}^{n_p} \norm{\mathbf{c}^{(k)}}_2,
\label{eq:gatom}
\end{equation}
where $\mathbf{c}^{(k)}$ denotes the $k$-th row of $\mathbf{c}$. Note that the grouped atomic norm regularizer remains as a QP problem, which has better scalability than the nuclear norm. Another advantage is that there is only one hyperparameter $\gamma$ in this optimization problem. In this paper, hyperparameters are selected by cross validation with validation data $u_v(t),z_v(t)\in \mathbb{R}^{n_vP}$.

The algorithm for LTP system identification with grouped atomic norm regularization is summarized as follows.

\begin{algorithmic}[1]
 \STATE \textbf{Given} $n,P,u(t),z(t),u_v(t),z_v(t)$
 \STATE \textbf{Select} $N,\{w_k\},\gamma_{\text{grid}}$
 \STATE \textbf{Compute} $\mathbf{g}^a$
 \FOR{$\gamma=\gamma_{\text{grid}}$}
 \STATE $\mathbf{c}(\gamma)\leftarrow \text{arg}\ \underset{\mathbf{c}}{\text{min}}\ V_{LS}(\mathbf{g}^a \mathbf{c})+\gamma \cdot J_{GA}(\mathbf{c})\ \text{s.t.}\ (\ref{eq:st})$
 \STATE $\epsilon(\gamma)\leftarrow V_{LS}(\mathbf{g}^a \mathbf{c}(\gamma)|u_v(t),z_v(t))$
 \ENDFOR
 \STATE $\gamma^*\leftarrow \text{arg}\ \underset{\gamma}{\text{min}}\ \epsilon(\gamma)$
 \STATE $\mathbf{c}^*\leftarrow \mathbf{c}(\gamma^*)$
\end{algorithmic}

\textbf{Remark.} A similar grouping concept can be extended to multi-input and multi-output systems, where sub-models are defined as SISO FIR models for each input-output channel at each tag time. Similarly, the same atom in all these sub-models should have consistent sparsity and thus be grouped together.

\section{Numerical Results}
\label{sec:sim0}

In this section, the grouped atomic norm method is compared with other LTP system identification schemes. First, a simple physical system consisting of a variable-length pendulum is examined to show the effectiveness of the grouped atomic norm method in generating uniform low-order system models, in contrast to other low-order methods. This clearly points out the unique advantages of our proposed method. Furthermore, we demonstrate by Monte Carlo simulation that the proposed method, with additional sparsity constraints imposed, gives a better fitting to the system compared to other regularized methods. It also outperforms the subspace identification method under high noise levels.

The following five identification schemes for LTP systems are compared. The first four methods use the switched FIR model of order $N=100$. The least squares method (\textit{LS}) directly solves the minimization problem of function (\ref{eq:LS}) with respect to $\mathbf{g}$. The Hankel nuclear norm method (\textsl{Hank}) solves the least squares problem with the regularizer (\ref{eq:nuc}). The Hankel matrices are constructed with $m=20$. The atomic norm method (\textit{Atom}), which was proposed in Section~\ref{sec:atom}, solves problem (\ref{eq:atom}). Our proposed method, the grouped atomic norm method (\textit{GAtom}) modifies the problem (\ref{eq:atom}) with the grouped regularizer (\ref{eq:gatom}). The atom set used in \textit{Atom} and \textit{GAtom} is defined by the poles $w_k=r\cdot e^{j\phi}$, where $r=[0.02:0.02:0.98,0.99,0.999], \phi=[0:\pi/50:\pi]$ as suggested in \cite{Pillonetto_2016}. This gives a total of $n_p=2601$ poles. 

In addition to the methods formulated in previous sections, we also compare our method to the widely-used subspace identification method (\textit{Sub}). The method proposed in \cite{Verhaegen_1995} is applied, except that in singular value truncation, the empirical and manual step to determine the system order is replaced with cross validation over a uniform order grid between 2 to 10. This automates the process for the Monte Carlo study.

The optimization problems are parsed by CVX and solved by MOSEK. In terms of computational time, \textit{LS} and \textit{Sub} are the fastest with closed-form solutions of unconstrained least squares problems; \textit{Hank} is slower than \textit{Atom} and \textit{GAtom} because of its SDP nature. The difference is more significant as period length $P$ and data length $nP$ increase.

\subsection{Case Study}
\label{sec:sim}

\begin{figure}[htbp]
\centerline{\includegraphics[width=1in]{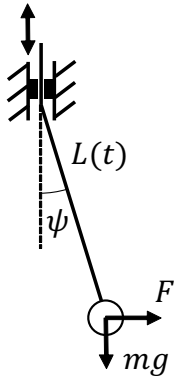}}
\caption{Illustration of the variable-length pendulum system.}
\label{fig:pen}
\end{figure}

Consider a pendulum of variable length shown in Fig.~\ref{fig:pen} with a periodic length profile $L(t)=L_0+l\cos\omega t$. The non-linear dynamics of the system are given by
\begin{equation}
\ddot{\psi} = -\frac{g}{L(t)}\sin\psi + \frac{2\omega l\sin\omega t}{L(t)}\dot{\psi}+\frac{1}{mL(t)}F\cos\psi,
\end{equation}
where $g$ is the gravitational acceleration. The parameters used are listed as follows.
\begin{equation*}
L_0=10\text{ m},l=5\text{ m},m=5\text{ kg},g=9.8\text{ m/s}^2,\omega=4\pi\text{ rad/s}
\end{equation*}
We model this system as a discrete-time SISO LTP system at small $\psi$, with $F$ as the input and $\psi$ as the output. The period length $P$ is selected as 4 with a sampling time of $T_s=2\pi/(P\omega)$. A data set of length $nP=500$ is simulated with a unit Gaussian input $u(t)\sim N(0,1)$ and output noise of $\sigma^2=(0.1\pi/180)^2$ for identification.

\begin{figure}[htbp]
\centering
\subfloat[\textit{Hank}]{\includegraphics[width=3.45in]{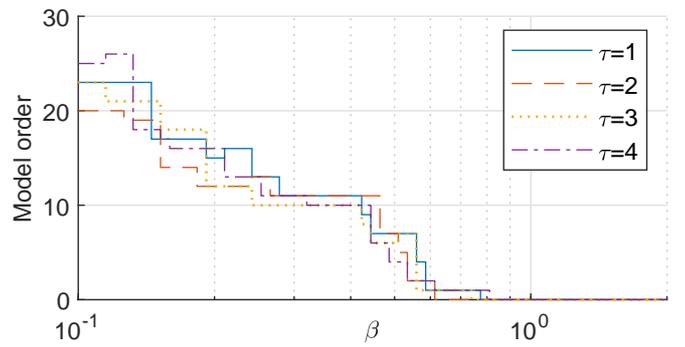}}\\
\subfloat[\textit{Atom}]{\includegraphics[width=3.45in]{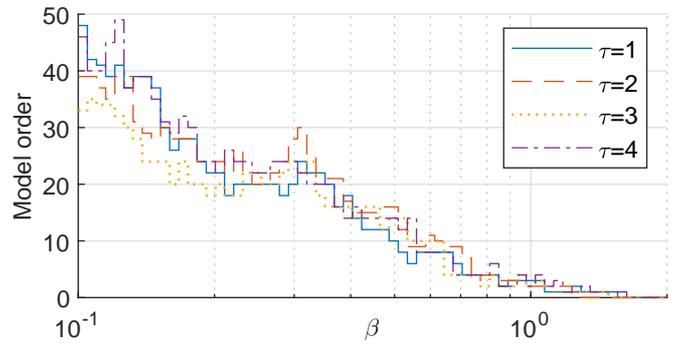}}
\caption{Estimated sub-model orders with sub-model complexity tuning.}
\label{fig:order1}
\end{figure}

\begin{figure}[htbp]
\centerline{\includegraphics[width=3.45in]{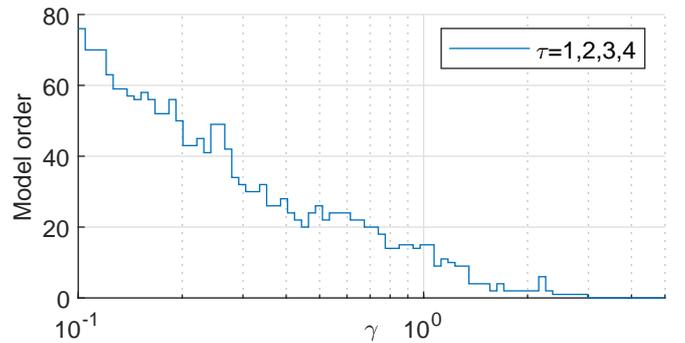}}
\caption{Estimated sub-model orders with \textit{GAtom}.}
\label{fig:order2}
\end{figure}

First, we try to obtain a uniform low-order model by fine tuning of sub-model complexity coefficients $\beta$. In this example, $\beta$ is selected from a 100-point log-space grid between $10^{-1}$ and $10^1$, and $\gamma$ is fixed to 1. The relations between the $\beta$ values and the estimated model orders are shown in Fig.~\ref{fig:order1}. It can be seen that since the model order is indirectly controlled by weighting parameters with no explicit expression, the sub-models cannot be regularized to any given order for both \textit{Hank} and \textit{Atom}. This makes it hard to tune the sub-model orders to be uniform, especially as $P$ increases. In contrast, \textit{GAtom} always gives a uniform estimation for any choice of the scalar hyperparameter $\gamma$ with the same grid, as shown in Fig.~\ref{fig:order2}. These uniform models can then be selected by cross validation. Note that in this paper, for the atomic methods, atoms from a fixed grid of poles are adopted which may not match the true pole locations. So multiple atoms may be selected to describe one pole. Active pole grid refinement is needed if the true order of the system needs to be recovered, which is out of the scope of this paper.

\subsection{Monte Carlo Study}

To compare the fitting performance of the proposed method with other methods, a Monte Carlo test campaign is set up as follows.

\subsubsection{System bank} A bank of 100 low-order discrete-time SISO LTP systems of period length $P=2$ is generated. The system orders are randomly selected between 2 and 10. Dynamics at each tag time $\{A(\tau),B(\tau),C(\tau)\}$ are generated by the {\sc MATLAB} function \texttt{rss}. These continuous-time systems are sampled at 3 times their bandwidths and discretized by zero-order hold equivalence. They are also normalized to have a DC gain of 1. The resulting LTP systems are verified to be stable.

\subsubsection{Data set} The systems are excited by Gaussian inputs with a unit variance, $u(t)\sim N(0,1)$. The outputs are perturbed at two different output noise levels, $\sigma^2=0.1$ and $0.01$. The initial states of the systems are set to 0. Two data sets of length $nP=500$ are generated for identification and validation respectively.

\subsubsection{Performance metric} The performance of the estimators is assessed by comparing to the true model with the following fitting metric
\begin{equation}
W=100\cdot \left(1-\left[\frac{\sum_{\tau=1}^P\sum_{i=1}^{100}(g^{\tau}_i-\hat{g}^{\tau}_i)^2}{\sum_{\tau=1}^P\sum_{i=1}^{100}(g^{\tau}_i-\bar{g})^2}\right]^{1/2}\right),
\end{equation}
where $g^{\tau}_i$ are the true impulse response coefficients in model (\ref{eq:smod}), $\hat{g}^{\tau}_i$ are the estimated coefficients, $\bar{g}$ is the mean of true coefficients. This metric extends that used for the \texttt{compare} function in the System Identification Toolbox to LTP systems. The state-space model obtained by \textit{Sub} is transformed to impulse response coefficients by (\ref{eq:imp}) for performance comparison.

The hyperparameter $\gamma$ in the regularized methods is cross-validated over a 10-point log-space grid between $10^{-1}$ and $10^1$. The sub-model complexity is not tuned ($\beta_\tau=1$) for \textit{Hank} and \textit{Atom}, as this tuning is hard to be automated and often impractical to unify the orders of sub-models as can be seen from Section~\ref{sec:sim}.

\begin{figure}[htbp]
\centering
\subfloat[$\sigma^2=0.01$]{\includegraphics[width=3.45in]{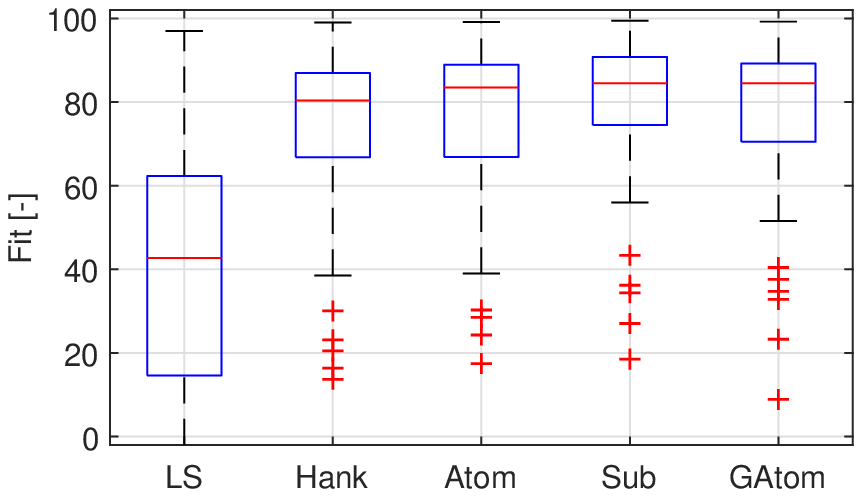}}\\
\subfloat[$\sigma^2=0.1$]{\includegraphics[width=3.45in]{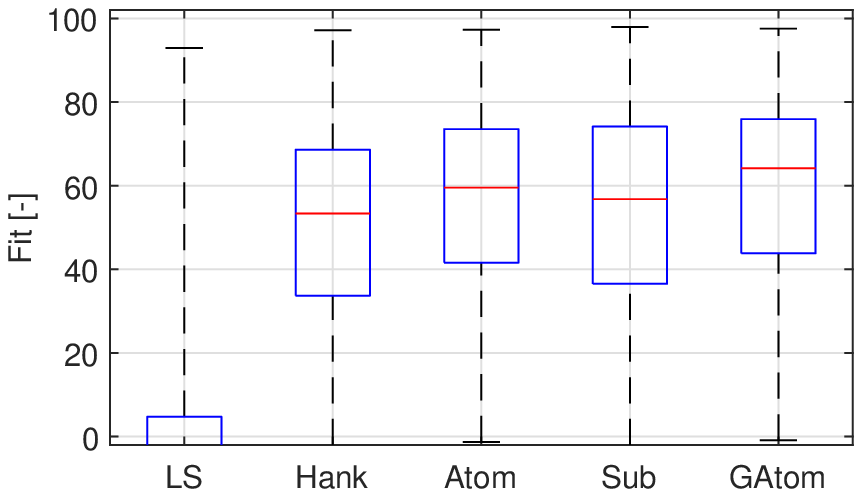}}
\caption{Comparison of fitting performance under different noise levels.}
\label{fig:mc}
\end{figure}

The results of Monte Carlo simulation are demonstrated by statistics in Table~\ref{tbl:mc} and box plots in Fig.~\ref{fig:mc}, under the low ($\sigma^2=0.01$) and the high ($\sigma^2=0.1$) noise levels respectively. It is shown that under both noise levels, the \textit{LS} method cannot give satisfactory estimates. Under the high noise level, the \textit{LS} estimation even fails to provide any information about the system with a negative average fitting. Comparing the three regularized methods, our proposed \textit{GAtom} method achieves the best model fitting by incorporating the requirement on pole locations. \textit{Atom} performs better than \textit{Hank} due to its guaranteed stability.

\textit{Sub} has an advantage over \textit{GAtom} under the low noise level with a higher mean fitting and a lower standard deviation. This is due to the fact that the subspace identification without regularization is a consistent estimator that converges to the true value in the noise-free case, whereas regularized methods are in general inconsistent. However, the advantage of \textit{GAtom} in model fitting is demonstrated under the high noise level, which is of more interest under realistic testing conditions.

\begin{table*}[htbp]
\renewcommand{\arraystretch}{1.2}
\caption{Statistics of Fitting Performance.}
\label{tbl:mc}
\centerline{
\begin{tabular}{cccccc|ccccc}
\hline\hline
                & \multicolumn{5}{c|}{\textbf{$\sigma^2=0.01$}}                               & \multicolumn{5}{c}{\textbf{$\sigma^2=0.1$}}                                 \\ \cline{2-11} 
                & \textbf{LS} & \textbf{Hank} & \textbf{Atom} & \textbf{Sub} & \textbf{GAtom} & \textbf{LS} & \textbf{Hank} & \textbf{Atom} & \textbf{Sub} & \textbf{GAtom} \\ \hline
\textbf{Mean}   & 21.4        & 68.6          & 70.6          & 78.8         & 71.7           & -106.1      & 47.8          & 52.6          & 48.7         & 55.6           \\
\textbf{Median} & 42.7        & 80.4          & 83.5          & 84.5         & 84.5           & -79.5       & 53.3          & 59.5          & 56.8         & 64.2           \\
\textbf{Std}    & 86.2        & 37.8          & 38.5          & 22.6         & 39.9           & 203.5       & 36.5          & 36.2          & 48.3         & 34.1           \\ \hline\hline
\end{tabular}
}
\end{table*}

\section{Conclusions}
\label{sec:con}

In this paper, we have proposed a new LTP system identification method with grouped atomic norm regularization. This method uses decomposed LTI sub-models to reformulate LTP systems for identification. A key requirement for the identification to be successful is that the sub-models should have the same pole locations. Therefore, the regularizer extends the atomic norm regularizer for LTI system to LTP systems with the group lasso technique to impose this additional structure. This method obtains uniform low-order models of LTP systems and simulations show it to have a better model fit compared to existing methods under high noise levels.

The main message of this work is that the LTP system identification problem cannot be fully tackled by LTI system theory. The key to enhancing the performance of LTP system identification is to incorporate specific structural constraints arising from periodicity with appropriate frameworks.

\bibliography{refs}             

\end{document}